\documentclass[bibnotes,twocolumn,aps,pra,a4paper]{revtex4}

\usepackage{color,epsfig,rotating,graphicx,subfigure}
\usepackage{amsmath,amssymb,amscd,ulem}

\usepackage{dsfont}
\usepackage{textcomp}

\newcommand{\kb}{k_{\rm B}}

\begin{document}

\title{Reply to: Reply to "On the heat transfer across a vacuum gap mediated by Casimir force"}

\date{\today}

\author{Svend-Age Biehs$^{1,*}$, Achim Kittel$^1$, and Philippe Ben-Abdallah$^2$}
\affiliation{$^1$ Institut f{\"u}r Physik, Carl von Ossietzky Universit{\"a}t, D-26111 Oldenburg, Germany}
\affiliation{$^2$ Laboratoire Charles Fabry, UMR 8501, Institut d’Optique, CNRS, Universit\'{e} Paris-Saclay, 2 Avenue Augustin Fresnel, 91127 Palaiseau Cedex, France}
\email{ s.age.biehs@uni-oldenburg.de} 

%%%%%%%%%%%%%%%%%%% abstract and OCIS codes %%%%%%%%%%%%%%%%

%\begin{abstract}
%\end{abstract}

\maketitle
The publication of a paper in Nature \cite{FongEtAl} and some followup comments made by the authors in different feature articles published in \cite{PhysicsToday,SpaceHeaters} urged us to clarify some statements on the ground of basic physical principals. We published this clarification in \cite{SABetalZNA2020} by picking up the original findings and putting them in the correct physical context. In the comment \cite{LiEtAl2020} to the preprint version \cite{SABetalArxiv2020} of our article~\cite{SABetalZNA2020} the authors claim that ``S.-A. Biehs et al. interpret that the Casimir heat transfer effect was not observed in our experiment'' which refers to the experiment in \cite{FongEtAl}. The authors of the comment \cite{LiEtAl2020} raise three points why S.-A. Biehs et al. are wrong in their opinion: because of ``(1) a misleading comparison between thermal radiation of bulk solids and Casimir force induced heat transfer through single phonon modes, (2) a fallacious interpretation of the heat flux measurement in our work[3], and (3) misunderstanding of the concept of mode temperature''. 

Before we start to reply to the points (1)-(3) we want to emphasize that the presentation in \cite{FongEtAl} is confusing and together with statements in that paper \cite{FongEtAl} and the statements of some of the authors accessible in the internet lead to not only confusing but simply wrong receptions of that work and its implications as in more popular publications like~\cite{PhysicsToday,SpaceHeaters}, for instance. This was our main motivation to discuss this experiment on Casimir force driven heat flux (CFDHF) in~\cite{FongEtAl} in a critical and general fashion in~\cite{SABetalZNA2020}. Since the analysis of the experiment in \cite{FongEtAl} uses the classical limit of the same model which we have used namely that in Ref.~\cite{SAB}, all formal expressions are of course the same in both works. 

Now let us reply to point (1). In our article~\cite{SABetalZNA2020} we compare heat radiation between the two membranes used in the experiment \cite{FongEtAl} with the CFDHF. It is argued in the comment \cite{LiEtAl2020} that this is misleading because ``Comparing these two distinct energy scales in a single plot [...] is misleading. It is like comparing the energy scales of heating up a glass of water and heating up a single water molecule.'' 
Actually, we have made this comparison because in  \cite{FongEtAl} it is claimed that the CFDHF ``has practical implications to thermal management in nanometer-scale technologies'' and that the experiment ``paves the way for the exploitation of quantum vacuum in energy transport at the nanoscale''. In more popular statements in the internet it is then further claimed by the authors of \cite{FongEtAl} that `` `For example, in hard drives, the magnetic read/write head moves above the disk surface with a separation as little as three nanometer,' Zhang says. `At such a short distance, the new heat-transfer effect is expected to play a significant role and so should be considered in the design of magnetic recording devices.' ''~\cite{SpaceHeaters}. And based on such statements it is further claimed that ``Phonon heat transfer through vacuum could have implications for managing heat in integrated circuits. The mechanism may provide a new way to intentionally dissipate heat in high-density transistor circuits, and it is an important consideration for keeping close elements thermally isolated in, for example, optical communication devices, which are sensitive to temperature cross talk.''~\cite{PhysicsToday}. We think these selling arguments go much too far. They are simply not relevant and even wrong. The authors never showed at what point or under which circumstances the CFDHF can contribute to the total heat transfer in a non-negligible way to justify statements like that. To emphasis this we illustrated the situation in the experiment by plotting the radiative heat transfer together with the CFDHF pointing out clearly that there is no reason to believe CFDHF plays any role in any practical application, especially under the circumstances of the experiment. This comparison shows that heat radiation is 15 orders of magnitude larger than the CFDHF. We further compared the CFDHF to the heat flux through a single molecule which is still 11 orders of magnitude larger~\cite{SABetalZNA2020}. This comparison is not misleading but rather it clarifies the role the CFDHF is likely to play on the proposed applications. Hence, it makes clear that to cool down a solid of  area $A$ using the CFDHF it is necessary to have at least $10^{15}$ strongly coupled resonant oscillator modes to be more efficient than with heat radiation or to have a unattainable number of modes. A similar comparison applies to the single molecule where the area $A$ would be the cross section of the molecule and it would be necessary to have $10^{11}$ single modes to beat a single molecule. We want to emphasize that predictions on the phonon tunneling or Van der Waals/Casimir driven heat flux for macroscopic objects (i.e. not single modes but bulk plates) might beat heat radiation but only at very small distances below 5\,nm~\cite{Budaev} or even much below 1\,nm~\cite{Pendry,Joulain,GangChen}. In all of these cases, at distances considered in the reported experiment the VdW/Casimir driven heat flux is again many orders of magnitude smaller than the radiative heat flux. For example, in Ref.~\cite{Pendry} it is shown in Table I that at a distance of 10\,nm the different models give for two gold plates a VdW/Casimir driven heat flux of $5 \times 10^{-5}$ or $7\times 10^{-7} \,{\rm W}/{\rm m}^2$ whereas they find a radiative heat flux of about $10^{6}\,{\rm W}/{\rm m}^2$ in Fig.~10(a) which makes a difference of 11 to 13 orders of magnitude.

Now, we turn to (2). In our article \cite{SABetalZNA2020} we discuss that for the determination of heat flux the expression 
\begin{equation}
  P_{A \rightarrow B} = 2 \gamma_A \kb (T_A - T_A')
\end{equation} 
(see  Eq. (16) in \cite{SABetalZNA2020}) is used, where $\gamma_A$ stands for the membrane damping rate, $T_A$  the membrane temperature and $T_A'$ the mode temperature which depends on the coupling constant of the two membranes. Our criticism is that the experiment in \cite{FongEtAl} is using this formula which is taken from the oscillator model (which we believe is correct no matter if one oscillator coupled to an external force or two oscillators are considered) to determine the heat flux and therefore is no direct heat flux measurement. This is done in the experiment by measuring the mode temperature $T_A'$ by measuring $\langle x^2_A \rangle$ (using the equation for  $P_{A \rightarrow B}$ from the model for one oscillator coupled to a bath and an external force), but could also be done by measuring only $g$ (which determines $\langle x^2_A \rangle$ or the mode temperature $T_A'$ in the coupled oscillator model as can be seen in Eq.~(2) below). So that equipped with this formula and the oscillator model any Casimir force measurement for membranes at two different temperatures could give a value for the heat flux.   
 
Our point is that in a heat flux measurement which deserves this name an independent heat flux measurement has to be carried out by means of a calibrated heat flux sensor. Here this would be some calibration by which a relation between an added heat flux to the single oscillator and the mode temperature change is made experimentally which would be an experimental proof of Eq.~(1). Such a calibration is not made but Eq.~(1) is assumed to be valid without any further confirmation. This is our only criticism concerning the performed heat flux measurement. Therefore we argue that $\langle x^2_A \rangle$ or the mode temperature (which are equivalent) are the directly measured quantity and the values of the heat flux can only be obtained indirectly with Eq.~(1). If an independent calibration measurement would have been carried out first, we would have nothing left to criticize here. We do not think that this argumentation is ``scientifically incorrect'' or a ``fallacious interpretation of the heat flux measurement''. Furthermore, our statement is correct that ``without the equation for $P_{A \rightarrow B}$ the experiment could not give any number for the Casimir force driven heat flux''~\cite{LiEtAl2020,SABetalArxiv2020}.  

Finally, we reply to point (3). The first claim is that we confuse mode temperatures and coupling constants
because in~\cite{SABetalArxiv2020} we wrote ``In the experiment $\langle \hat{x}^2 \rangle$ has been measured which is of course a mean to determine $g$''. We make a similar statement in \cite{SABetalZNA2020}. The authors of the comment~\cite{LiEtAl2020} conclude that ``This  statement  is incorrect. Mode temperature (proportional  to  mean  squared displacement) and coupling  constant  are  two  distinct physical  quantities that  can  be  measured independently''~\cite{LiEtAl2020}. First let us make clear that we are not saying that the mode temperature and coupling constant cannot be measured independently. But as we detail in our article~\cite{SABetalZNA2020} the oscillator model on which the analysis of the experiments relies on directly relates the mean square displacement $\langle x^2_A \rangle$ or mode temperature $T_A'$ and the coupling constant $g$ as can be seen in Eq.~(13) in \cite{SABetalZNA2020} which reads
\begin{equation}
	  T_A' = T_A + \frac{g^2 \gamma_B (T_B - T_A)}{(\gamma_A + \gamma_B)(g^2 + \gamma_A \gamma_B)}.
\end{equation}
In the experiment the damping rates of both membranes $\gamma_A$ and $\gamma_B$, the membrane temperatures $T_A$ and $T_B$ are known quantities so that by a measurement of the mode temperature $T_A'$ or $\langle x^2_A \rangle$ one can determine $g$ (and vice versa) which would be a measurement provided that this relation is verified experimentally. Actually Fig.~3(a) of the experiment~\cite{FongEtAl} shows the measurement of the left hand side of this relation in comparison to the theoretically evaluated right-hand side. Hence this mutual dependence of the mode temperature and the coupling constant is experimentally verified and is to our understanding the main result of the experiment~\cite{FongEtAl}. As a side comment, this relation could also be verified by using any other forces coupling both membranes with its specific distance dependence for sure. Furthermore, by plugging the experimentally measured left hand side for $T_A'$ and the theoretically evaluated right hand side into Eq.~(1) one can directly convert it in a heat flux without measuring any heat flux which gives Fig.~3(e) in Ref.~\cite{FongEtAl}. In other words, our statement is correct and it is astonishing that it is questioned, because~\cite{FongEtAl} have proofed it experimentally by proofing the validity of Eq.~(2). 

Then the authors of the comment~\cite{LiEtAl2020} further claim that we do not understand the mode temperature concept: ``They even make a general claim that `in the strong coupling limit  the  widely  used  concept  of  mode  temperatures loses  its  thermodynamic  foundation  and therefore cannot be employed to make a valid statement on cooling and heating'~\cite{SABetalArxiv2020}. Such claims are again incorrect.  As we discussed above, mode temperature directly quantifies the energy of the phonon mode and its deviation from bath temperature provides information of heat flux. It is valid regardless of the coupling strength, and it is valid when the oscillator is not in thermal equilibrium with the thermal bath.''~\cite{LiEtAl2020}. As a reply, we can only emphasize that our statement is that the mode temperature is not a thermodynamic temperature in the strong coupling regime where the membranes are decoupled from their baths. We are not questioning the validity of the definition of the mode temperature as claimed. We define the mode temperature in~\cite{SABetalArxiv2020,SABetalZNA2020} of course in the same way as in~\cite{FongEtAl} so that one can hardly speak of a misunderstanding on our side. What we say is, that a statement on cooling or heating (as needed in the above suggested applications) can only be made when a thermodynamic temperature is changed and the mode temperature is not a thermodynamic temperature in the strong coupling regime. The mode temperature is a concept which cannot be used to deal with cooling/heating mechanisms if the mode is not a part of a heat bath. Or loosely speaking when holding a thermometer on one of the strongly coupled membranes this will definitely not measure the mode temperature.

To illustrate the difference between the mode temperature and the thermodynamic temperature let us consider the following simple example. It is a weak motor which drives a pendulum via a spring with mass $10\:{\rm kg}$ such that the mass of the pendulum reaches a height of $h=1\:{\rm m}$. In this case, the energy content of the pendulum is $E=98\,{\rm J}$ and according to the equipartition theorem  this oscillator’s energy corresponds to a temperature $T=2 k^{-1}_B E$ that is to $T\approx1.4 \times 10^{25}\:{\rm K}$ . Obviously this 'mode' temperature has nothing to do with a thermodynamic temperature. Otherwise the system as a whole would immediately evaporate. Furthermore, it is not correct to speak about a heat transfer from the motor to the pendulum or a cooling of the motor.  

Hence, speaking of mode temperatures might be misleading for practical applications because they can be confused with and sometime assimilated to thermodynamic temperatures in particular when it is not always specified that one is speaking about a mode temperature. This clearly is what happened in Fong et al.'s paper in Fig. 3a which leads to confusion as in \cite{PhysicsToday} where obviously the mode temperature is understood as the thermodynamic membrane temperature. 

We hope that our response to the comment~\cite{LiEtAl2020} makes clear that the reproaches that we are "misunderstanding" basic concepts, giving "fallacious" misinterpretation of the measurement of the heat flux, and "misleading" comparisons between the radiative heat flux and the CFDHF cannot be upheld. A careful reading of our critical work~\cite{SABetalZNA2020} will confirm that our arguments are not ``scientifically incorrect'', but scientific and worth consideration. Our criticism of the mode temperature concept is not questioning its validity in general but only its interpretation in terms of a thermodynamic temperature in the case where the mode is decoupled from any heat bath and the possible confusion caused by speaking of temperatures and cooling/heating. The analysis of the experiment could have been presented without using the notion of mode temperatures at all by using only mean square displacements or occupation numbers. Furthermore, the heat flux measurement is done without a calibrated heat flux sensing setup and therefore definitely hinges on the oscillator model (no matter if only one or two oscillators are considered). Therefore it is in our opinion not a direct and independent heat flux measurement. Finally, our comparison of the radiative heat flux and the CFDHF (and the heat flux through a single molecule and the CFDHF) are legitimate and show that this CFDHF is extremely small and can be neglected at distances larger then 10\,nm in comparison to radiative heat transfer. Misleading are rather the proposed ideas for applications of this effect and ideas like ``In principle, stars may even heat their planets through this newfound mechanism. Given the distances involved, however, the magnitude of this effect would be `exceedingly small,' essentially to the point of utter insignificance, Zhang says.''~\cite{SpaceHeaters} The mechanism of CFDHF does not play an important role at 10\,nm, at 1\,\textmu m, at 1\,mm, and it is definitely nonsense to believe it is worth a single word of a scientist at a distance of 1\,AU.

\end{document}